\documentclass[showpacs,aps,prb,twocolumn,superscriptaddress]{revtex4-1}

\usepackage{graphicx}
\usepackage{dcolumn}
\usepackage{bm}

\usepackage{color}
\usepackage{amsmath}

\usepackage[FIGTOPCAP,raggedright,normalsize,tight]{subfigure}

\begin{document}

\title{Ripple edge engineering of graphene nanoribbons}

\author{Philipp Wagner}
 \email{philipp.wagner@cnrs-imn.fr}
\author{Christopher P. Ewels}
 \email{chris.ewels@cnrs-imn.fr}
\author{Viktoria V. Ivanovskaya}
 \affiliation{Institut des Mat\'eriaux Jean Rouxel (IMN), Universit\'e de Nantes, CNRS UMR 6502, 44322 Nantes, France}
\author{Patrick R. Briddon}
\affiliation{School of Electrical, Electronic and Computer Engineering, University of Newcastle, Newcastle upon Tyne, NE 1 7RU, United Kingdom}

\author{Amand Pateau}
\author{Bernard Humbert} 
\affiliation{Institut des Mat\'eriaux Jean Rouxel (IMN), Universit\'e de Nantes, CNRS UMR 6502, 44322 Nantes, France}

\date{\today}

\begin{abstract} 

It is now possible to produce graphene nanoribbons (GNRs) with atomically defined widths. GNRs offer many opportunities for electronic devices and composites, if it is possible to establish the link between edge structure and functionalisation, and resultant GNR properties. 
Switching hydrogen edge termination to larger more complex functional groups such as hydroxyls or thiols induces strain at the ribbon edge. However we show that this strain is then relieved via the formation of static out-of-plane ripples.
The resultant ribbons have a significantly reduced Young's Modulus which varies as a function of ribbon width, modified band gaps, as well as heterogeneous chemical reactivity along the edge.
Rather than being the exception, such static edge ripples are likely on the majority of functionalized graphene ribbon edges. 
\end{abstract}

\pacs{81.05.ue, 62.20.de, 81.40.Jj, 31.15.E-, 81.05.uj}

\maketitle

\section{Introduction}

Since the start of the graphene boom in 2004 \cite{Novoselov2004}, the simplistic picture of a perfect flat carbon monolayer has been refined.  Notably dynamic rippling of graphene has been demonstrated via electron microscopy \cite{Meyer2007} and molecular dynamics calculations \cite{Fasolino2007}, and there have been first discussions of the possibility for stable static ripples in graphene \cite{Thompson-Flagg2009,Shenoy2008}.
In reality graphene is not an infinite plane but is constrained by edges. Graphene nanoribbons (GNRs)  have different properties from the infinite bulk material, notably they can display a finite band gap as a function of ribbon width \cite{Barone2006}.  In addition the physical and chemical behaviour of the one-dimensional edges will be superimposed on that of the bulk two-dimensional graphene.  Graphene nanoribbons can be produced by unzipping carbon nanotubes \cite{Kosynkin2009}, lithography \cite{Tapaszto2008,Fasoli2009}, etching \cite{Bai2009} and controlled chemical bottum-up methods \cite{Cai2010}, and offer great potential both for nanoelectronics \cite{Chen2007} and nanocomposites \cite{Su2009}.  Edges are an easily accessible way to chemically functionalise the graphene and hence modify its properties. We show here that careful design of graphene edges allows us to define the graphene nanoribbon properties.

The simplest way to saturate GNR edge dangling bonds is via hydrogen termination, which is the standard approach in GNR modelling \cite{Nakada1996,Wassmann2008}.  However other terminating heteroatoms can be imagined (e.g. N, O), and an intriguing example of this was a recent theoretical study which found F-terminated armchair GNRs to be more stable when twisted helically \cite{Gunlycke2010}.\\ 
However to develop a more realistic picture of the possibilities of GNR edge chemistry, more complex termination groups have to be investigated. A good example are hydroxyl groups (OH), which as well as being bulkier than simple heteroatoms also show more complex chemical interaction between themselves.  Such groups have been proposed as a way to introduce strain along the graphene edge \cite{Peng2011}, in order to tune GNR electronic properties such as bandgap \cite{Gui2008,Guinea2010}.  
However two dimensional layered materials such as graphene have alternative mechanisms for relieving edge-induced strain, namely structural deformation into the third dimension via rippling or buckling \cite{Shenoy2008}.
This additional degree of freedom adds significant richness to graphene edge chemistry, which we investigate here for infinitely long armchair graphene nanoribbons (AGNRs) using the example of hydroxyl functionalisation.
In the current study we show how -OH termination of different width AGNRs modifies their structure, strain and stability. We then demonstrate the effect of such functionalisation on band gap, chemical reactivity to metal deposition, and Young's Modulus, and generalise to other functional groups and ribbon types.
This study points the way towards ``edge termination engineering" as a way to create GNRs with custom designed properties.\\

\section{Method}

We performed density functional theory calculations under the local density approximation as implemented in the AIMPRO code \cite{AIMPRO, AIMPRO2}. The calculations were carried out using supercells, fitting the charge density to plane waves within an energy cut-off of 200 Ha. Electronic level occupation was obtained using a Fermi occupation function with $kT = 0.04$~eV. Relativistic pseudo-potentials are generated using the Hartwingster-Goedecker-H\"utter scheme \cite{HGH}. These functions are labelled by multiple orbital symbols, where each symbol represents a Gaussian function multiplied by polynomial functions including all angular momenta up to maxima $p$ ($l$ = 0, 1) and $d$ ($l$ = 0, 1, 2). Following this nomenclature, the basis sets used for each atom type were $pdddp$ (C), $ppp$ (H) and $dddd$ (O), resulting in 38 independent functions for carbon, 12 for hydrogen and 40 for oxygen.  A more detailed account of the basis functions can be found elsewhere \cite{Goss2007}. A Bloch sum of these functions is performed over the lattice vectors to satisfy the periodic boundary conditions of the supercell. 

\begin{figure}
\includegraphics[height=5cm]{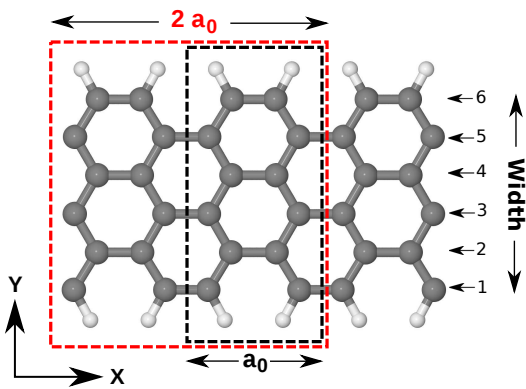} 
 \caption{(Color online) H-terminated AGNR, unit cells shown by dotted frames, definition of armchair ribbon width shown by numbers \cite{CervantesSodi2008}, width 6. C atoms are shown grey and H atoms are white.}
 \label{H_ribbon}
\end{figure}

Supercell sizes have been checked and chosen to be sufficiently large (y- and z-distance between ribbons $> 12$ \AA) to avoid interaction with neighbouring GNRs. A fine k-point grid was chosen of the form  $12 /(n \cdot a_0) \times 1 \times 1$ with $n \in N$ where $a_0$ is the length of a fundamental unit cell along the ribbon axis in the supercell (see Fig.\ref{H_ribbon}), 
which gives energies converged to better than 10$^{-5}$~Ha. For rippled GNRs the unitcell was doubled along the ribbon ($2 \cdot a_0$) to satisfy the periodic conditions of the supercells (longer period ripples using $n \cdot a_0$ cells, $n > 2$ were also tested but found to be less stable) . For all structures the atom positions and the lattice parameters have been fully relaxed. The definition of ribbon widths of AGNRs is given by Cervantes-Sodi et al.
\cite{CervantesSodi2008} and is used in the text and Fig.\ref{4to20energy} and Fig.\ref{OH_gap_difference} (definition see Fig.\ref{H_ribbon}). 

\section{Results and discussion}

\begin{figure}
\subfigure [][] {\includegraphics[height=2.5cm]{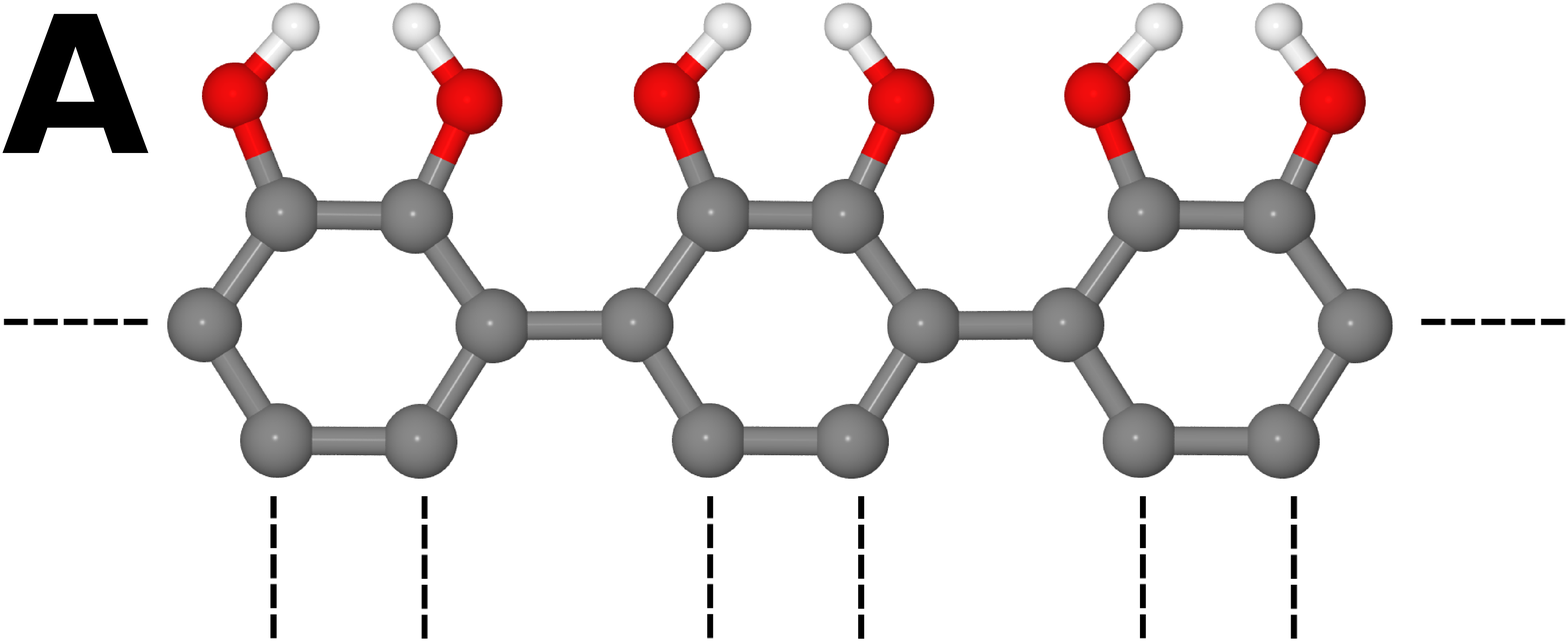}} \\
\vspace{0.3cm}
\subfigure [][] {\includegraphics[height=2.5cm]{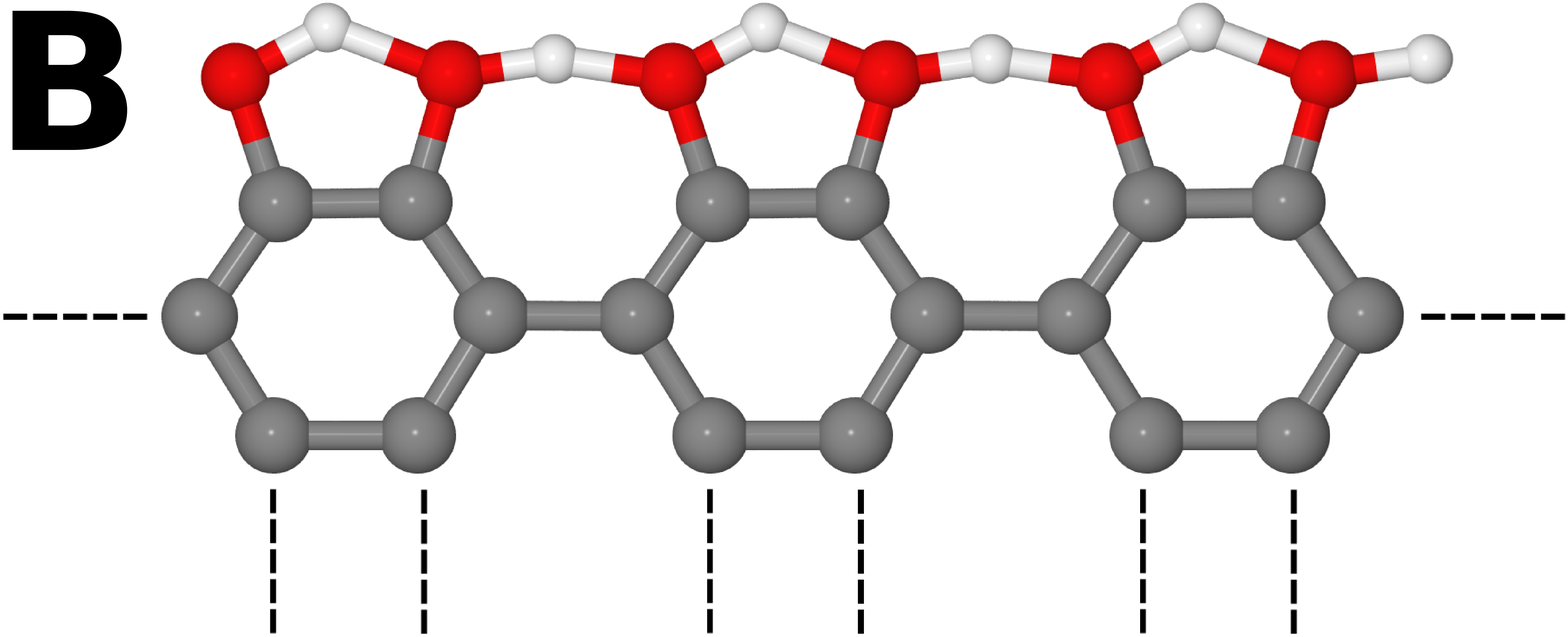}} \\
\vspace{0.3cm}
\subfigure [][] {\includegraphics[height=2.9cm]{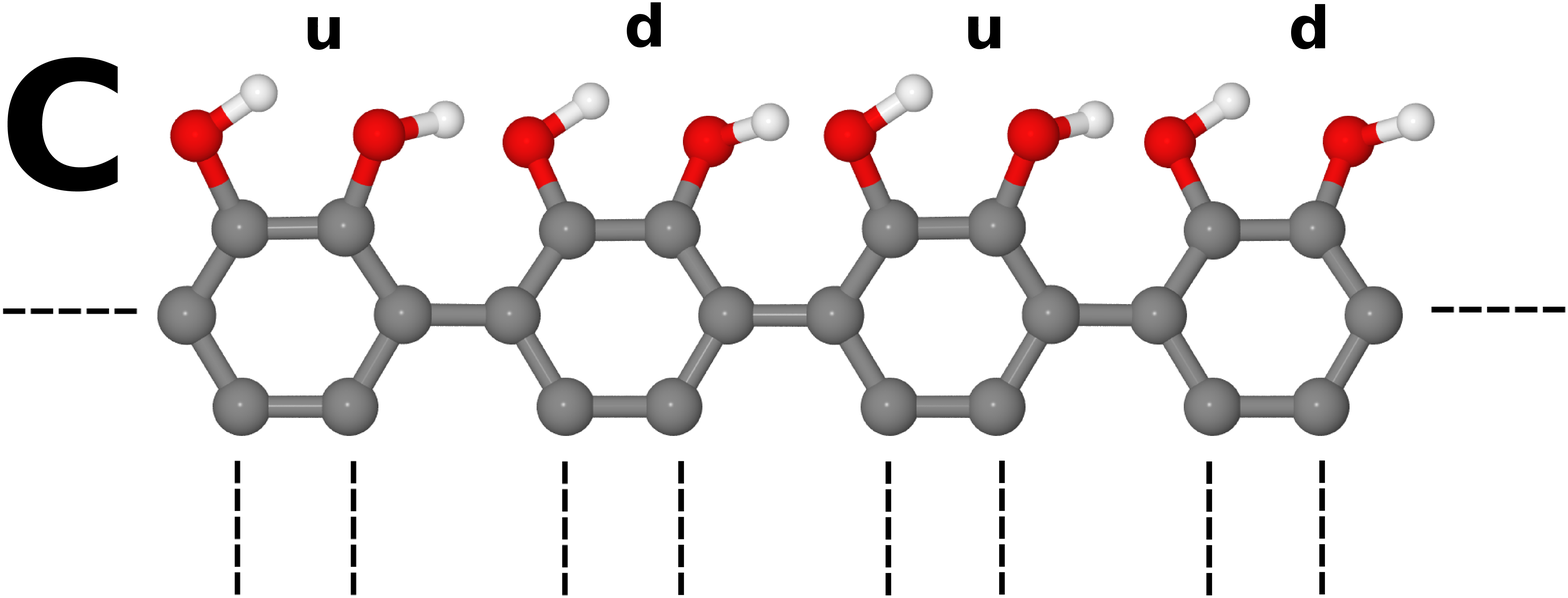}} \\ 
\vspace{0.3cm}
\subfigure [][] { \includegraphics[width=7cm]{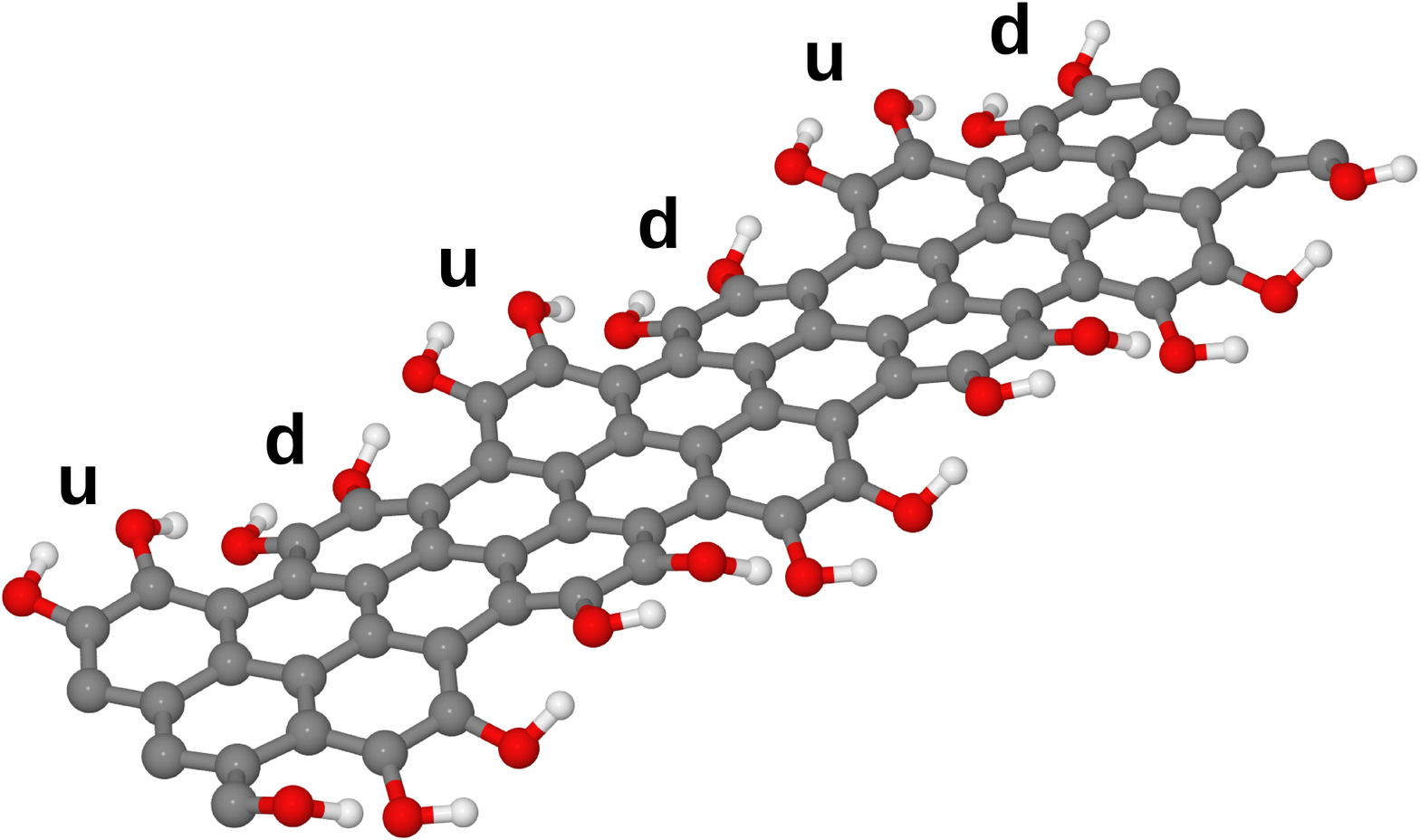}} \\
\vspace{0.3cm}
\subfigure [][] { \includegraphics[width=7cm]{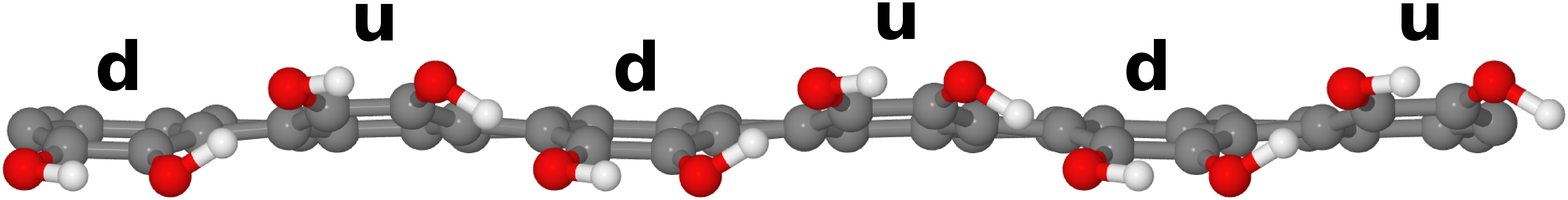}}

 \caption{(Color online) (a)-(c): three OH-terminated armchair edge structures: A (metastable base symmetric structure), B (metastable with hydrogen bonding) and C (stable ground state structure with static rippling). (d)-(e): stable OH-terminated AGNR  of type C (width 6), (d) perspective view of ribbon and (e) side view with clearly visible rippled edge. Dotted lines are a guide to the eye for the GNR. $u$ stands for ``up" and $d$ for ``down" for the rippled edge. C atoms are in grey, O atoms in red and H atoms are white (same for Fig.\ref{OH_uudd}, Fig.\ref{Pd-addition} and Fig.\ref{chiral}).}
 \label{OH_flat_good_bad}
\end{figure}

Larger functional groups on the GNR edge can introduce a range of possible inter-group interactions, including steric hindrance, Coulombic repulsion, dipole-dipole interactions and hydrogen bonding.  Taking the case of hydroxyl groups we attempt to address these systematically.  
We have modelled fifteen different structural possibilities but focus here on three structures of hydroxylated edges for GNRs with widths from 4 to 20 (refered to hereafter as structures A, B and C). The relevant relaxed edge structures are shown in Fig.\ref{OH_flat_good_bad} and associated structural parameters in Table \ref{table_bl}. 

\begin{table}
\begin{tabular}{lr|l |l| l }
    \hline
    \multicolumn{2}{l|}{Structure}& $d_{C-O}$ (\AA) & $d_{O-H}$ (\AA) & $d_{O\cdots H}$ (\AA) \\
    \hline
    \hline
    \multicolumn{2}{l|}{A (flat)} & $1.33 $ & $0.96 $  & - \\
    \hline
    B (flat) &width 4& $1.36 $ & $1.05$ & $1.26$/$1.41 $ \\
             &width 20& $1.35$ & $1.09$ & $1.11$/$1.14$ \\
    \hline
    C (rippled) & width 4 & $1.36 $ & $1.02 $ & $1.40 $/$1.63$ \\
     						& width 20 & $1.36 $ & $1.02 $ & $1.45 $/$1.69$ \\
    \hline
    \multicolumn{2}{l|}{Phenol ($\rm C_6H_6O$)} & $1.35$ & $0.97$ & - \\
    \hline   
  \end{tabular}
  \caption{Overview of calculated bond lengths of the AGNRs from width $4-20$ for structures A, B and C. O$\cdots$H bond lengths for Structure B and C vary with ribbon width between the limits given in the table.  Structure C is the stable ground state. Calculated Phenol values included for comparison.}
 \label{table_bl}
\end{table}

\begin{figure*}
\subfigure [] { \includegraphics[width=8.5cm]{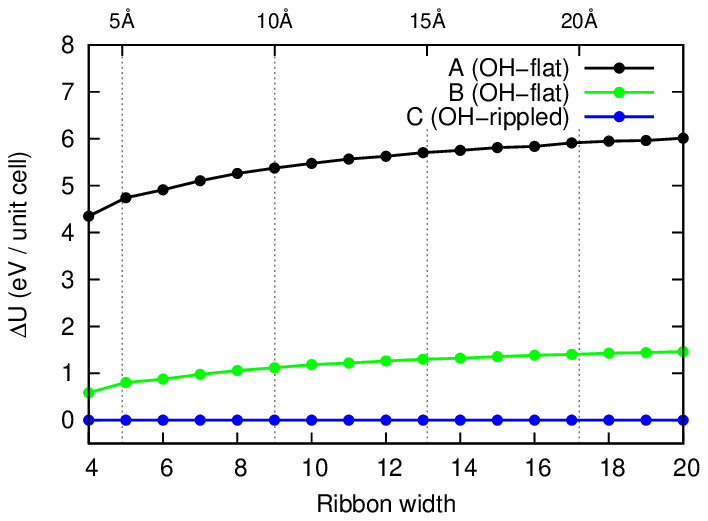}}
\subfigure [] { \includegraphics[width=8.5cm]{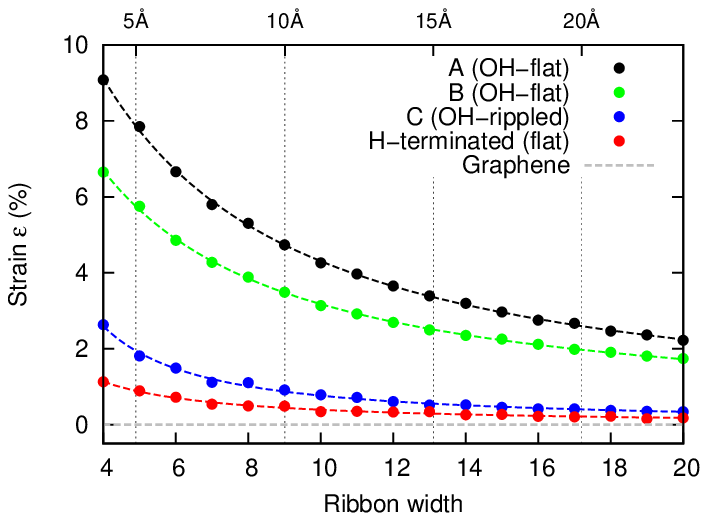}}
 \caption{(Color online) (a): Energy difference $\Delta U$ between structure A (black) and structure B (green) referring to structure C (blue) versus ribbon width from 4 to 20.  (b): Strain along the ribbon (x-axis see Fig.\ref{H_ribbon}), strain is referenced to a perfect flat graphene sheet ($\epsilon = 0$ \%), H-termination (red), A (black), B (green) and C (blue). Fits with $\epsilon(x)=\frac{a}{x^2} + \frac{b}{x} +c$ (fit parameter in Supplementary Materials\cite{supmat}). The top scale of (a) and (b) in \AA \, is taken from perfect flat graphene with C-C bondlengths of $d_{c-c}= 1.41$ \AA \, as guide for the eye (also for Fig.\ref{OH_gap_difference}).}
 \label{4to20energy}
\end{figure*}

We start with structure A, symmetry constrained to lie in the plane with symmetrically paired -OH groups, removing any possibility for hydrogen bonding.
Calculated bond lengths are $d_{C-O}=1.33$ \AA \, and $d_{O-H}=0.96$ \AA. 
This structure was recently discussed as a possible configuration for inducing strain in AGNRs \cite{Peng2011}. 
Indeed we find that it induces strains $\epsilon$ of up to 9 \% for small AGNRs (see Fig.\ref{4to20energy},b). However this structure is extremely unstable. Strain along the ribbon has been calculated as follows (see also Fig.\ref{H_ribbon}):
$$\epsilon = \frac{a_0-a_{Graphene}}{a_{Graphene}}$$ 
with $a_{Graphene} = 4.23 \;\mbox{\AA}$ our optimized DFT (LDA)
lattice parameter along the armchair direction of free standing graphene, in good agreement with experimental values of $\approx 4.2 \;\mbox{\AA}$\cite{Grass2008}.

We next introduce hydrogen bonding between the hydroxyl groups by breaking the in-plane symmetry, giving the most stable planar configuration, structure B.
Here $d_{C-O}=1.36$ \AA \, with hydrogen forming one strong covalent bond $d_{O-H}=1.05-1.09$ \AA \,
and one weaker hydrogen bond $d_{O \cdots H}=1.11-1.41$ \AA. This new hydrogen bond lowers the system energy by typically 3.7~eV/unit cell compared to Structure A, and reduces
slightly the induced strain (still up to $6.7 \%$ for small AGNR compared to perfect graphene, as can be seen in Fig.\ref{4to20energy},b).

However by additionally breaking the planar symmetry we reach the energetically most stable configuration for hydroxyl terminated AGNRs, structure C.  
The -OH groups displace out of plane pairwise, creating a static sinosoidal ripple along the AGNR edge (see Fig.\ref{OH_flat_good_bad},d,e). 
By displacing out of plane the structure releases up to $1.46$ eV/unitcell (Fig.\ref{4to20energy},a) and the strain in the ribbon is
relieved (Fig.\ref{4to20energy},b), returning to values similar to hydrogen terminated AGNRs.  The carbon pairs for each hexagonal ring are displaced
up and down alternately (longer period oscillations tested in supercells with $n \cdot a_0$ and $ n > 2$ at a variety of ribbon widths were found to be less stable).  
While $d_{C-O}$ stays largely unchanged at 1.36 \AA, the covalent $d_{O-H}$ extends to $1.02$ \AA, while the rippling allows the hydrogen bond lengths to increase to $d_{O\cdots H}=1.40-1.69$ \AA.
These lengths are approaching our calculated bondlengths for phenol ($d_{C-O}=1.35$ \AA, $d_{O-H}=0.97$ \AA, Table \ref{table_bl}).

\begin{figure}
\centering
\subfigure [][] { \includegraphics[width=4.4cm]{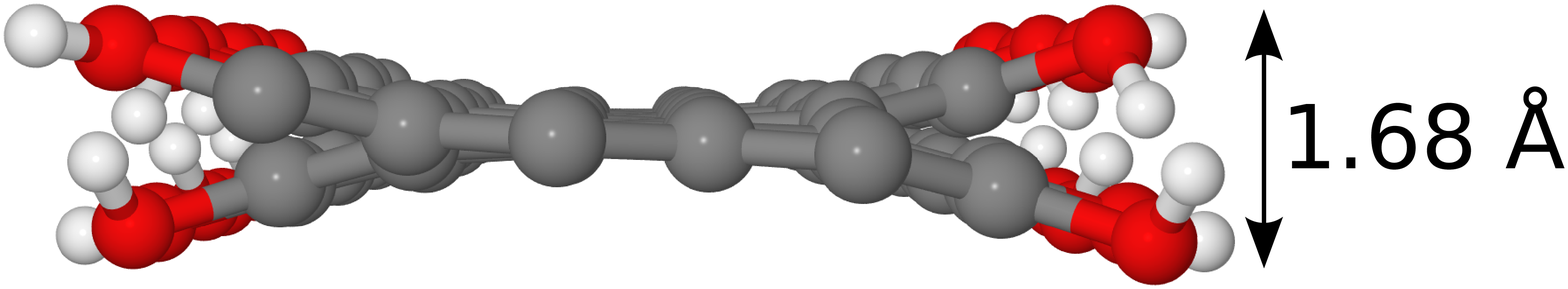}}\\
\subfigure [][] { \includegraphics[width=7cm]{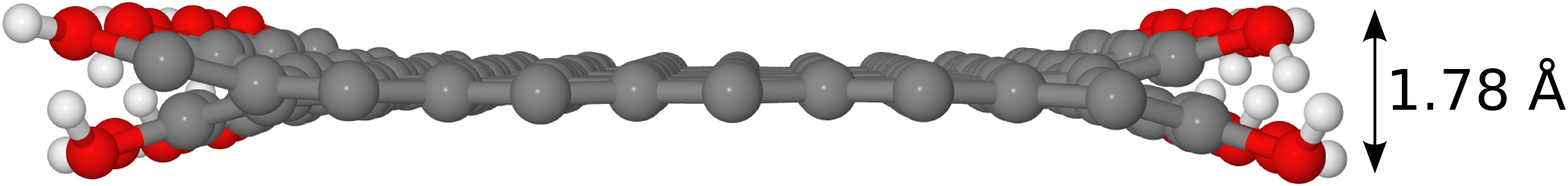}}\\
\subfigure [][] { \includegraphics[width=8.5cm]{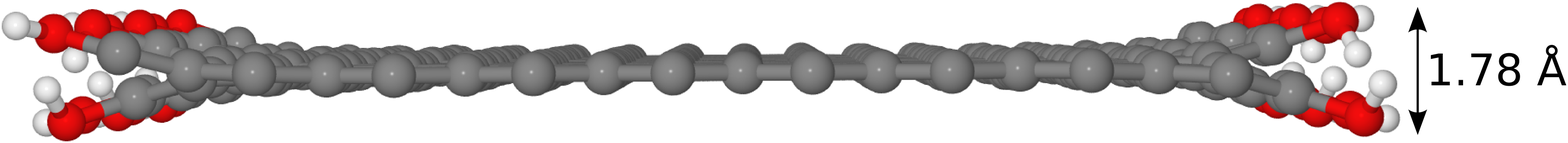}}
\caption{(Color online) Front view of -OH terminated AGNRs (stable structure C), (a) width 6, (b) width 12 and (c) width 18.}
\label{OH_uudd}
\end{figure}

The out of plane deformation mode can be understood as an elastic response to a 1D edge line tension applied to a rigid 2D graphene sheet (our calculated Young's Modulus for graphene is 1.08 TPa), where it is clearly energetically
favourable to buckle the edge rather than stretch the whole sheet.   However non-planar sheet deformation decreases the $\pi$-orbital overlap, and this enthalpic driving force of graphene to remain flat localises the rippling along the ribbon edge.  This can be seen in Fig.\ref{OH_uudd} where the ripple amplitude increases from $1.63/2 = 0.82$ \AA \, (width = 4) to $1.78/2 = 0.89$ \AA \, for large ribbons (width $\geq 12$), and where out-of-plane displacements remain localised within $\approx 2.3$ \AA \, of the graphene ribbon edge. The ripple amplitude is inversally proportional to the AGNR strain and only indirectly coupled to the width of the ribbon, consistent with a picture of a constant buckled edge length more able to dilate the graphene when its basal plane area is smaller.

No energy barrier was found between the rippled C and flat B structures, demonstrating that configuration B is a metastable maximum (see Supplementary Materials\cite{supmat}). The energy barrier calculations can be summed to estimate the ripple inversion barrier, giving a minimum barrier of $\Delta U_{Barrier} = 0.59$ $eV/a_0$ to invert one edge wave. This barrier is too high for thermal activation at room temperature, and structure C can be considered as a static edge rippling.

Thus we find that hydroxylated AGNRs will show spontaneous static edge ripples, due to an edge-strain induced out-of-plane deformation mode.  As a result the strain induced by the rippled hydroxylated edge is for wide ribbons less then $0.35$ \% (width $\geq 20$), and very close to that of H-terminated AGNRs of the same width ($< 0.18$ \%). We note that we found no significant interaction between the two edges of different OH-terminated AGNRs (as reflected in the total energy or strain), suggesting they are largely decoupled.\\

\begin{figure*}
\centering
\subfigure [][] {\includegraphics[width=8.5cm]{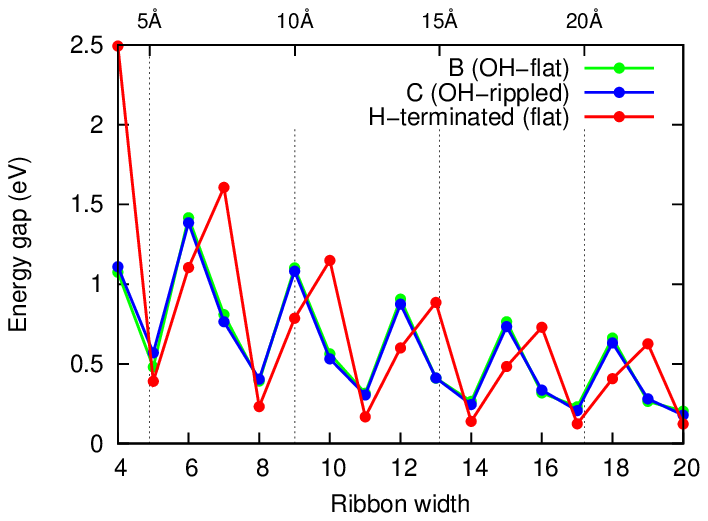}}
\subfigure [][] {\includegraphics[width=8.5cm]{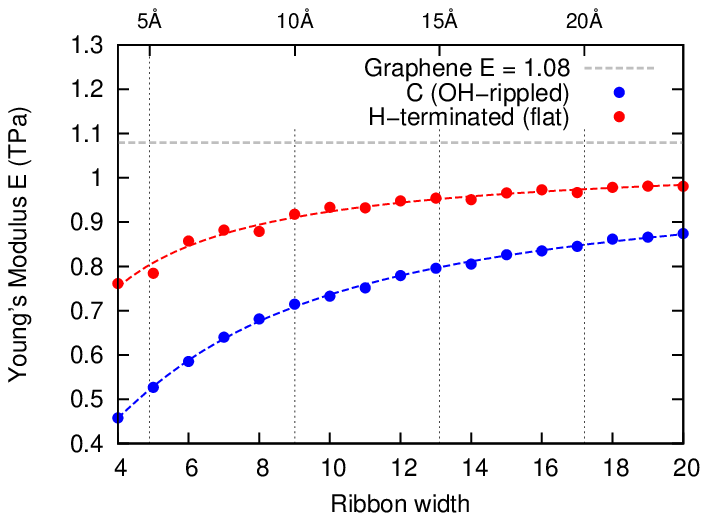}}
\caption{(Color online) (a): Energy gap versus width for different terminated AGNR. H-terminated AGNR (red), structure B (green) and structure C (blue). (b): Young's Modulus $E$ along the AGNRs (in case of graphene along the armchair direction). $E$ fitted with $E = \frac{a}{x} + \frac{b}{x^2} +c$ as guide for the eye (fit parameter in Supplementary Materials\cite{supmat}).}
\label{OH_gap_difference}
\end{figure*}

We next investigate how static ripples modify the ribbon properties, comparing structures B and C to determine the influence of rippling, and C and H-terminated ribbons to determine the influence of the hydroxyl functional group. Starting with electronic properties, we obtain good agreement between our calculated LDA band gap for flat H-terminated AGNRs and previous literature \cite{Barone2006}.  Bandgap is inversely proportional to ribbon width, superimposed with strong 3N periodicity.  This alternation is explainable via Clar sextet theory \cite{Baldoni2008} and Fermi wavelength \cite{Ezawa2006}.
The calculated band gap for structures B and C (see Fig.\ref{OH_gap_difference},a) are almost superposed, showing that edge rippling does not appear to affect the gap for -OH termination, which is instead dominated by the choice of edge functional group.  Once again we observe 3N periodicity \cite{Baldoni2008} but the difference with H-termination is not a simple phase shift of the periodicity.  For $3N+2$ ($N=1,2,..$) widths the bandgap is similar to the H-terminated case, but for $3N$ and $3N+1$ the difference is significant, with bandgap fluctuations of up to 50\%.
Thus ribbon band gap appears highly sensitive to choice of edge functional group, however the gap is not sensitive to out-of-plane edge rippling for -OH termination.

\begin{figure}
\centering
\subfigure [][]{\includegraphics[width=7cm]{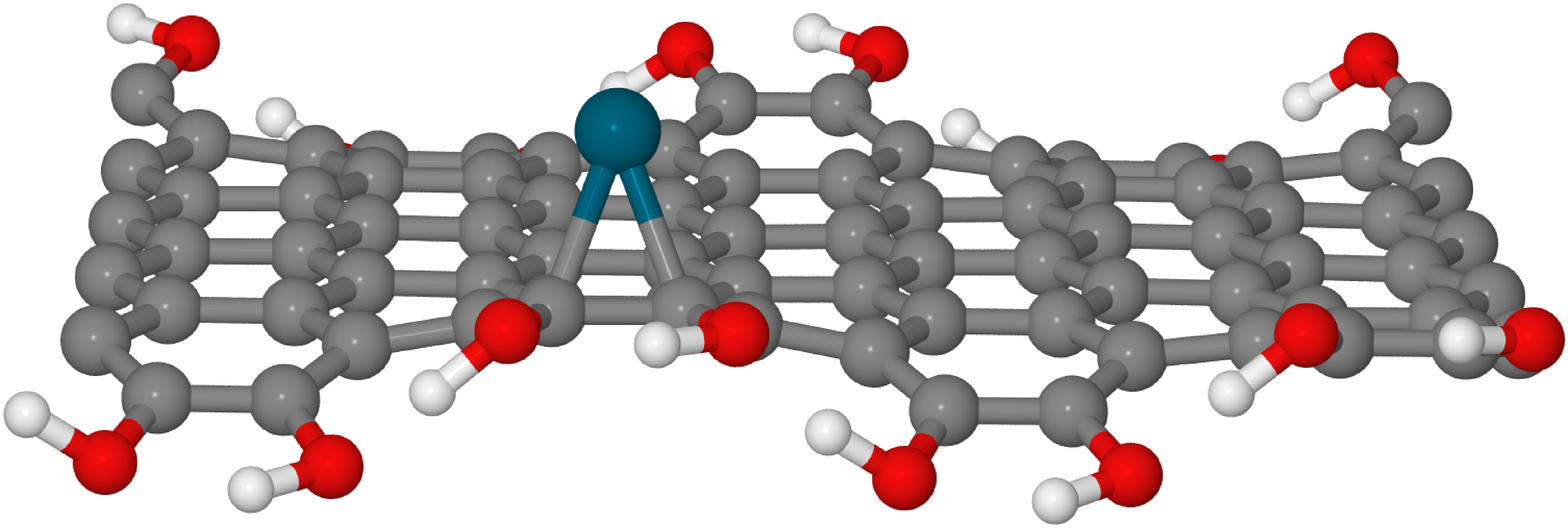}}
\subfigure [][]{\includegraphics[width=7cm]{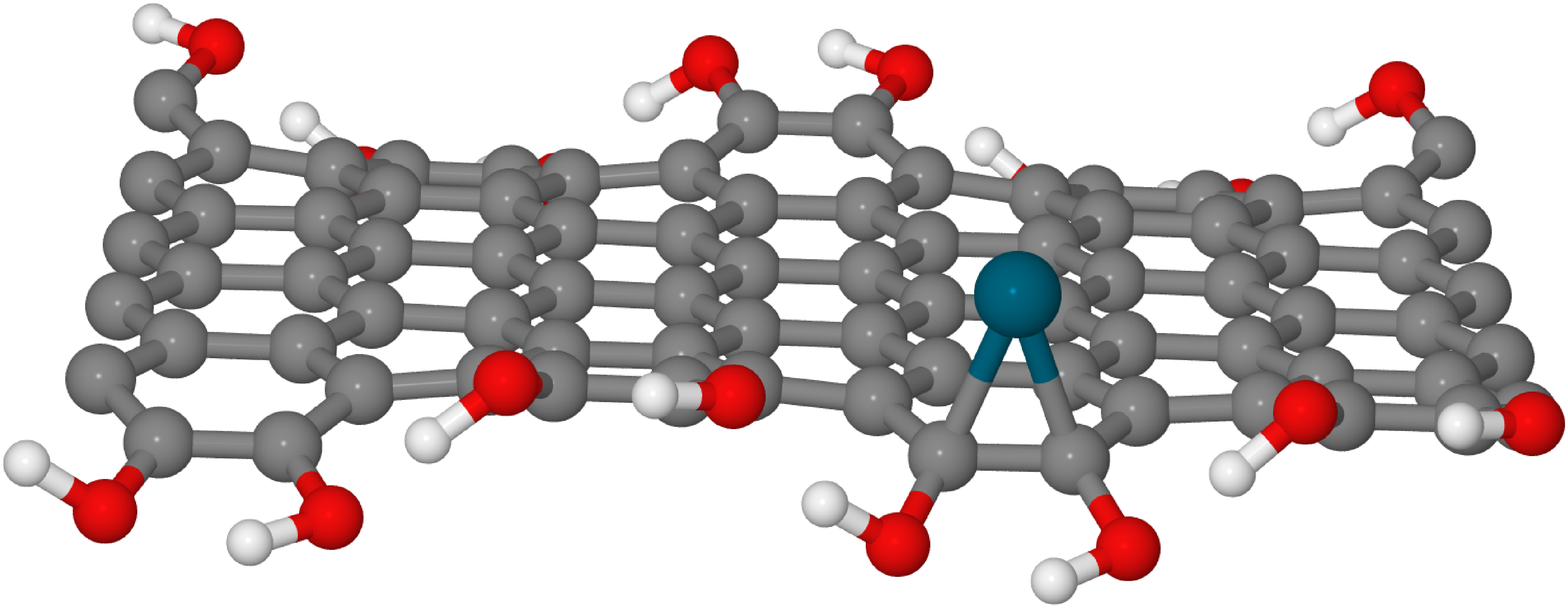}}

  \caption{(Color online) Pd atom addition to a -OH terminated AGNR (width 10). (a) ``Top-ridge" site, (b) ``Valley" site, which is $\Delta U = -0.4$ eV more stable.}
 \label{Pd-addition}
\end{figure}

Rippling may be expected to modify the chemical reactivity of the ribbon surface, and to estimate this we calculated possible bonding sites for single Pd atom addition on a AGNR (width 10).  Pd sits above C-C bond centres \cite{Suarez-Martinez2009,Lopez-Corral2010}, confirmed in our calculations.  In general binding nearer the ribbon edge is more stable. For rippled edges we find Pd atoms at the ribbon edge are 0.4 eV more stable in a concave ``valley" site of an edge ripple than on the convex ``ridge top" (see Fig.\ref{Pd-addition}). Indeed the concave ``top" site is even 0.1 eV less stable than a flat site at the ribbon centre. We expect that with enough activation energy the Pd atoms can migrate to the edge and will sit in the valleys of the rippled edge structure. Thus the ripple-induced changes in surface curvature periodically modify the surface reactivity of the graphene, suggesting interesting changes in absorption behaviour for chemisorbed and physisorbed species, and potentially important geometric effects for metal contact deposition.

The largest change due to edge rippling we find is in AGNR mechanical behaviour.
We determined Young's Modulus ($E$) of AGNRs along their length using the approach described by Zeinalipour-Yazdi et al \cite{Zeinalipour-Yazdi2009}. 
Inducing strain up to $\pm 3$ \% along the ribbon the total energies have been calculated (7-points).  Energy difference ($\Delta U$) vs. strain ($\epsilon$) was fitted with a quadratic function $\Delta U(\epsilon)=a\epsilon^2$. $E$ is given by $E = \frac{2 a}{V}$, where $V = a_0 \times w \times h$ is the volume of the ribbon section composed of length $a_0$, width $w$ taken as the H-H distance between the edges, and $h = 3.35$ \AA. 
We find $E$ for flat infinite graphene to be $1.08$ TPa, in good agreement with literature values ($1.09$ TPa\cite{Zeinalipour-Yazdi2009}, $1.05$ TPa\cite{Liu2007}). However finite width H-terminated and edge-rippled OH-terminated AGNRs show significantly smaller Youngs' Modulus (see Fig.\ref{OH_gap_difference},b).
In both cases modulus is approximately inversely proportional to ribbon width, with some non-linearity at smaller widths when edge effects start to dominate.  Both edge terminations extrapolate to the ideal graphene value at infinite width (H-terminated: $1.05$ TPa, OH-rippled: $1.03$ TPa), and  
as expected the modulus of -H terminated AGNRs drops towards the value for cis-polyacetylene at small widths. However surprisingly, changing the edge functionalisation from -H to rippled -OH groups significantly decreases the Youngs' Modulus of the ribbon.  This reduction is a direct result of the rippling (indeed calculations for the unstable flat -OH ribbons structure B, actually show a slight increase in Youngs' Modulus over the -H terminated case, since in this case the ribbon is under slight tension (see Supplementary Materials\cite{supmat})). We note that as for the band gap, a periodicity of $3N$ can be seen in the values of $E$ reflecting variations in C-C bonding with width.
These changes in $E$ as a function of both ribbon width, and edge termination, will be critical in graphene nanocomposite design.  Modifying ribbon edge functionalisation is an obvious way to chemically bind nanoribbons into a host polymer matrix, but these results show that changing the functionalisation groups can decrease the Young's Modulus of the nanoribbon itself as much as 40 \%.  

\begin{figure}
\centering
\subfigure [][]{\includegraphics[width=7.5cm]{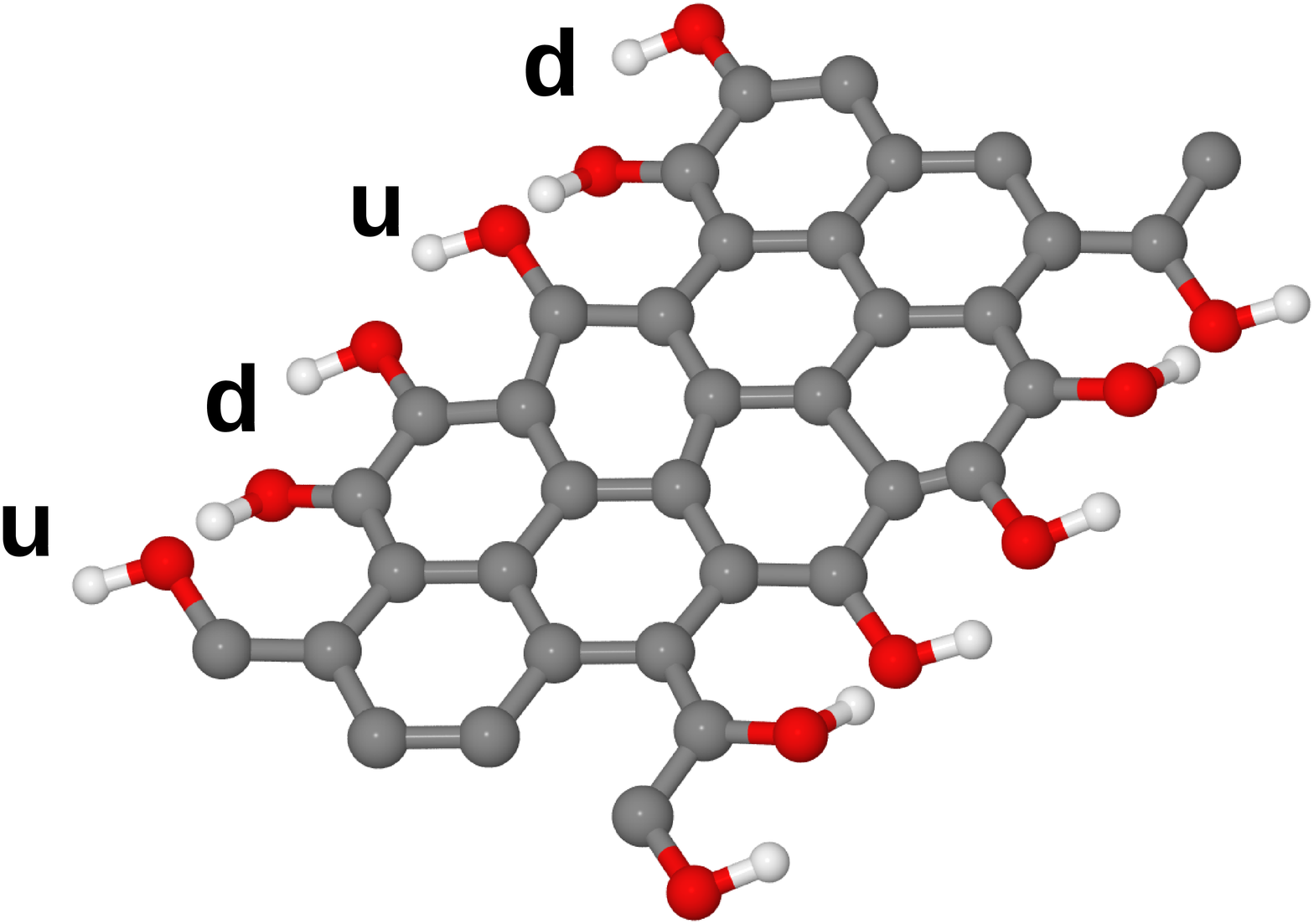}} \\
\vspace{0.6cm}
\subfigure [][]{\includegraphics[width=7.5cm]{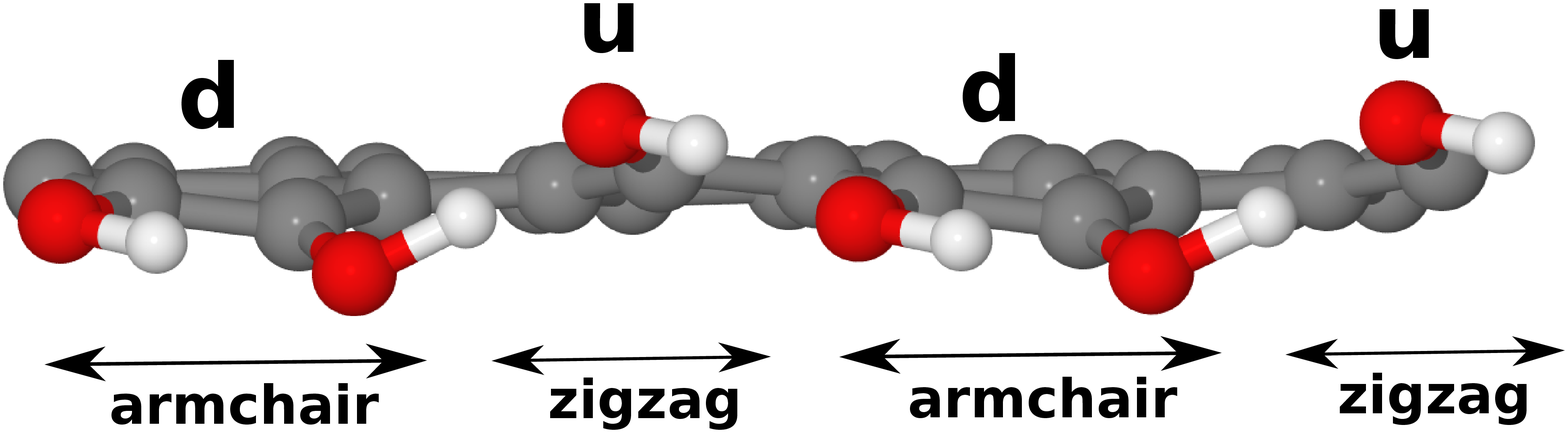}}\\
\vspace{0.6cm}
\subfigure [][]{\includegraphics[width=7.5cm]{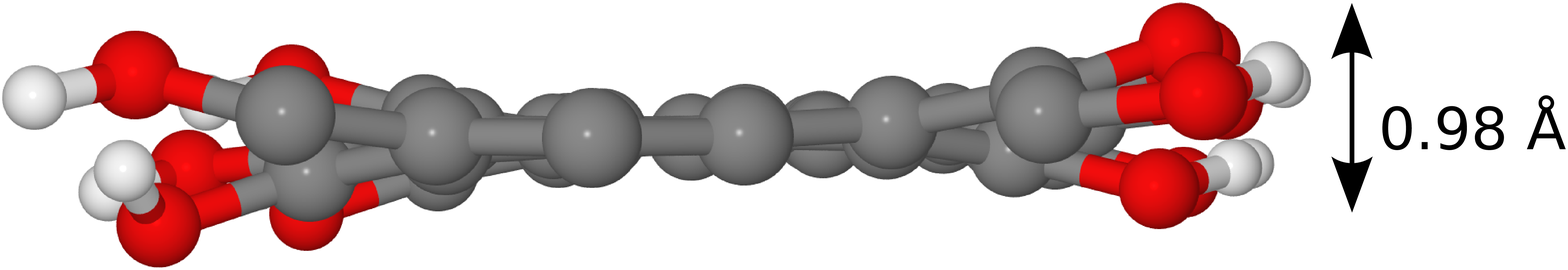}} 

  \caption{(Color online) Chiral -OH terminated GNR. (a) perspective view, (b) side view, (c) front view. 
  }
 \label{chiral}
\end{figure}

\begin{figure}
\centering
\subfigure [][]{\includegraphics[width=7cm]{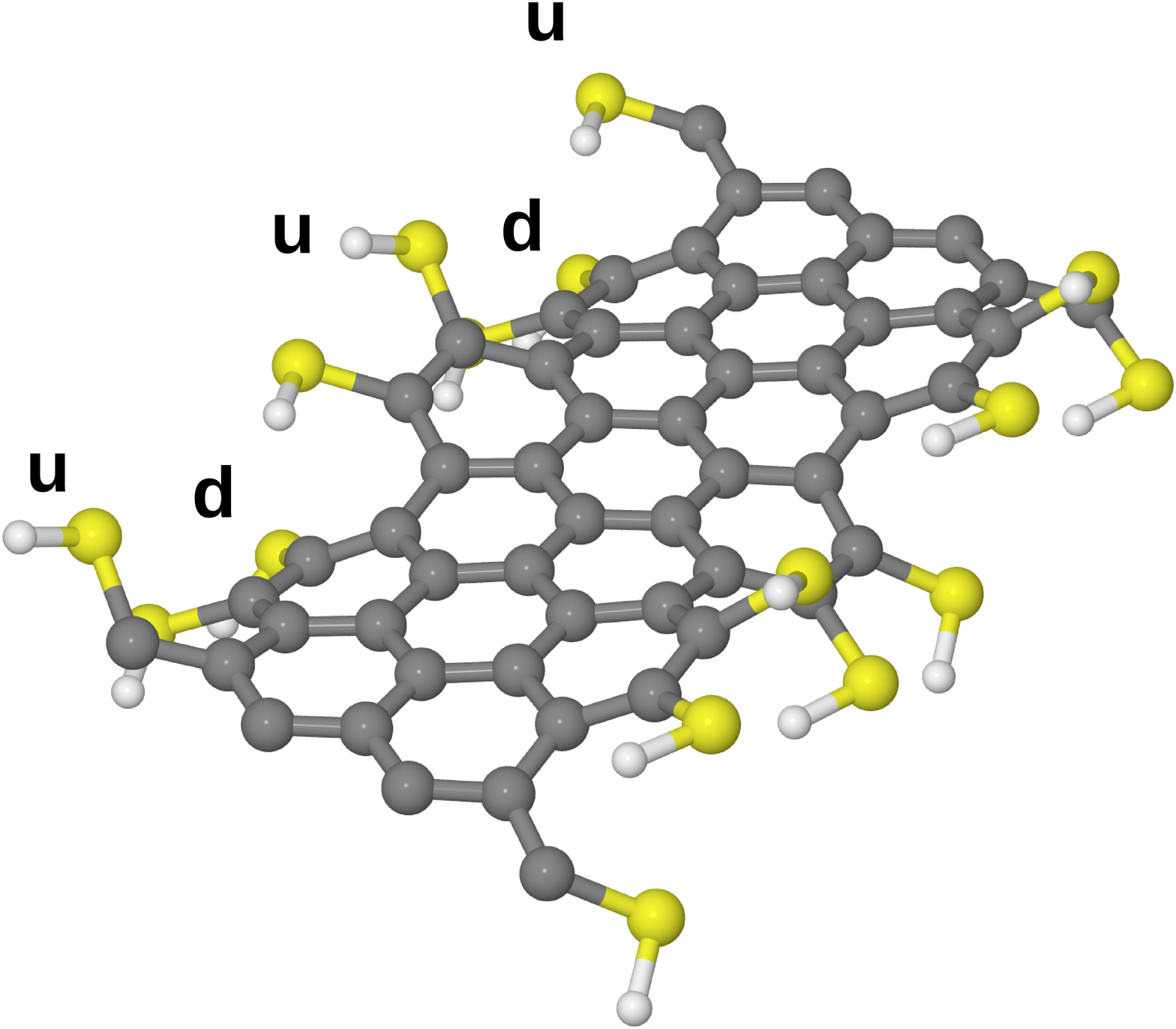}}\\
\vspace{0.3cm}
\subfigure [][]{\includegraphics[width=7cm]{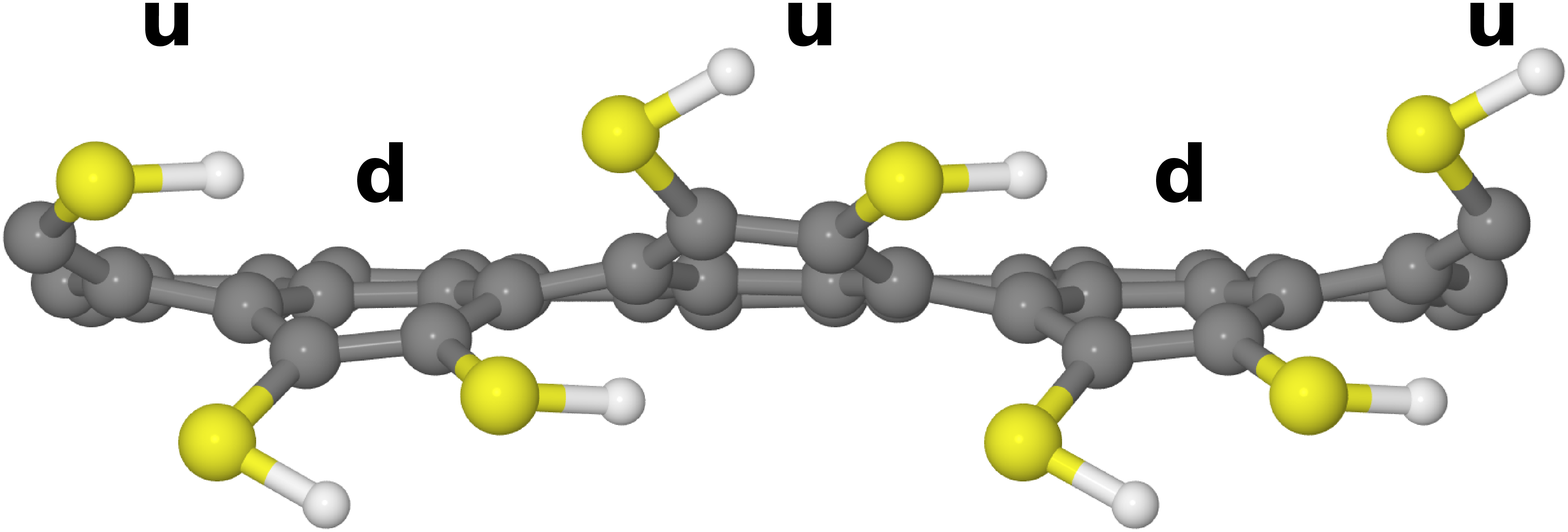}}\\
\vspace{0.3cm}
\subfigure [][]{\includegraphics[width=6cm]{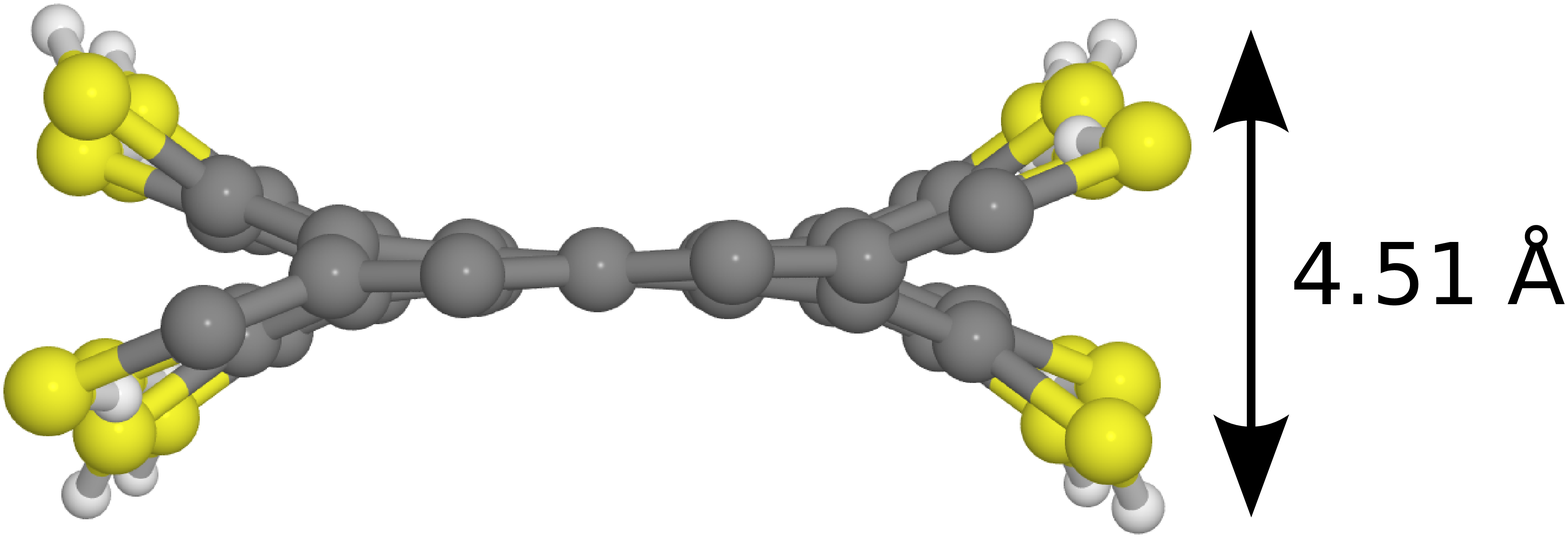}} 

  \caption{(Color online) AGNR (width 7) -SH terminated. (a) perspective view, (b) side view, (c) front view. In grey C atoms are pictured, H atoms in white and S atoms in yellow. 
 $u$ stands for ``up" and $d$ for ``down" for the rippled edge.
}
 \label{SH}
\end{figure}

In order to determine how general the static edge rippling effect is, we next examined other chiralities.
Hydroxylated zig-zag GNRs (ZGNRs) do not exhibit out-of-plane edge rippling, due to the larger spacing between edge C atoms (2.44 \AA) which results in sufficient spacing between the -OH groups to keep edge strain below $\pm 0.2$ \%. Indeed, in principle edge tension might be expected to induce rippling within the ribbon centre, the inverse effect of AGNRs, however the rigidity of the graphene sheet precludes this.  Note however this does not preclude rippling in ZGNRs for larger functional groups.  Calculations for hydroxylated chiral edges show an intermediate effect, with localised edge rippling around armchair-like sections which rapidly decays in zig-zag sections of the edge (see Fig.\ref{chiral}). Thus these results show that the majority of GNR edge types, when hydroxylated, will exhibit static rippling.  

We also examined the dependence of static edge rippling in AGNRs on functional group, replacing -OH with -F, -Cl, and -SH.  In all cases the ribbons exhibited the same periodic edge rippling as the hydroxylated edges (see Fig.\ref{SH} and Supplementary Materials\cite{supmat}).  Thus a key finding of this study is that flat GNR edges, as exemplified by hydrogen termination, appears to be the exception rather than the rule.  The majority of graphene nanoribbon chiralities and functionalisations we have examined undergo spontaneous out-of-plane static rippling.

\section{Conclusions}

This fundamental study provides a first picture of underlying physics for adding complex functional groups to graphene edges and in particular AGNRs.  Hydroxyl groups do not induce large strain in GNRs, since the strain is compensated by static ripple formation along the ribbon edge.  Such ripples form in the majority of ribbon chiralities and functional groups we have examined, suggesting flat ribbon edges as observed for hydrogenation may be the exception. This also seems to rule out edge functional groups as a simple way of inducing large strains in the graphene basal plane. For the first time we calculate Young's Modulus for infinite AGNRs with -H and -OH terminations, and show that edge ripples can drastically modify the mechanical properties and chemical reactivity of the ribbon.  We believe this could have a big impact on engineering devices and composites with embedded GNRs.   We find that the band gap 
is not sensitive to static edge rippling for -OH termination, but can change by up to 50 \% depending on choice of edge functional group.

There are various ways to match a 1D line tension along a ribbon edge against a 2D surface strain in the basal plane.  In the case of graphene, with a Young's Modulus of 1.08 TPa, the in-plane resistance to tension is high and edge compression is the energetically favoured solution, resulting in static ripple formation at the edge.  At the same time there is an energetic cost associated with rippling induced disruption of the graphene $\pi$-network, and hence the ripples remain localised near to the ribbon edge, reducing the effective flat basal plane width of the ribbon.  
In the case of chiral and aperiodic edges, static edge rippling is likely to lead to weak localisation and a decrease in the intravalley scattering length, consistent with experimental observations in graphene flakes \cite{Tikhonenko2008}.  

The discussion thus far concerns free-standing graphene.  On substrates the modulating height of the hydroxylated edge ripples will result in periodic modulation in the graphene-substrate spacing and hence the substrate-induced potential felt by the nanoribbon, increasing electronic localisation effects.  Variation in graphene-substrate spacing at the ribbon edge could also facilitate impurity intercalation beneath the graphene.
Finally we note that this static edge rippling behaviour is a fundamental response of a 2D layered system to 1D edge strain, and as such is also likely to also be of importance in the range of new monolayer materials under development such as BN, MoS$_2$ and NiTe$_2$ \cite{Coleman2011}.  Strain compensation through rippling is also likely to be important at other graphene interfaces such as grain boundaries.

\begin{acknowledgments}
PW, CPE and VVI thank the NANOSIM-GRAPHENE project n$^{\circ}$ANR-09-NANO-016-01 funded by the French National Agency (ANR) in the frame of its 2009 programme in Nanosciences, Nanotechnologies \& Nanosystems (P3N2009). We thank the CCIPL, where some of these calculations were performed. We thank COST Project MP0901 NanoTP for support.
\end{acknowledgments}



%

\end{document}